\documentclass[aps,pra,reprint,superscriptaddress,floatfix,longbibliography]{revtex4-2}

\usepackage{amssymb}
\usepackage{graphicx}
\usepackage{dcolumn}
\usepackage{bm}
\usepackage{amsmath}
\usepackage[colorlinks,linkcolor=magenta,citecolor=blue,urlcolor=blue]{hyperref}

\begin{document}
	
	\title{Dynamics of two particles with quasiperiodic long-range interactions}
	\author{Yun Zou}
	\email{12024117009@stu.ynu.edu.cn}
	\affiliation{School of Physics and Astronomy, Yunnan Key Laboratory for Quantum Information, Yunnan University, Kunming 650091, PR China}
	

	\begin{abstract}
		
		We investigate the dynamics of two identical spinless fermions on a one-dimensional lattice with open boundary conditions (OBC), subject to quasiperiodic long-range interactions. Using numerical exact diagonalization (ED), we study this non-integrable system as a continuous-time quantum walk and uncover a robust correlated dynamical regime. This regime, characterized by an approximately constant inter-particle distance, emerges under sufficiently strong quasiperiodic modulation of the long-range interactions. Further, the study shows that the behavior is determined by the nature of the interaction and the choice of boundary condition. Notably, by tuning the phase of the quasiperiodic modulation, we observe three distinct manifestations of this phenomenon: localization, nearest-neighbor separation oscillations, and next-nearest-neighbor separation transitions---each arising for specific initial separations. Furthermore, we identify the suppression of entanglement entropy in the system, including instances of oscillatory behavior. Our results highlight how quasiperiodic long-range interactions shape few-body quantum dynamics.
		
	\end{abstract}
	
	\maketitle
	
	\section{Introduction}
	Advances in quantum simulator technologies in recent years enable direct investigation of nonequilibrium dynamics in complex systems through precise Hamiltonian engineering. These quantum platforms allow continuous tuning of interaction parameters in the effective model, making the observation of resulting dynamics possible. Furthermore, their strong isolation from external noise supports long-lasting coherent dynamics in the simulated system~\cite{polkovnikov2011colloquium,altman2021quantum}. This has prompted an investigation into the dynamics of isolated complex systems---a topic of significant interest in theoretical physics, whose answer promises to boost developments in quantum technology~\cite{eisert2015quantum,hahn2024eigenstate}.
	
	The tunability of current quantum simulators allows for custom-designed interactions, fueling extensive research into the long-range interacting quantum systems~\cite{tran2020hierarchy,defenu2023long,mattes2025long}, which are more complex than short-range ones. Conventionally, “long-range” refers to interactions decaying as $1/r^\alpha$ with distance $r$ between particles in an interacting quantum system~\cite{maity2020one,guo2020signaling,defenu2023long,defenu2024out}. However, long-range interactions can also take other forms, such as the quasiperiodic modulation considered here. Long-range quantum systems are realized in various experimental platforms, such as Rydberg atom arrays~\cite{saffman2010quantum,bernien2017probing}, dipolar systems~\cite{carr2009cold,chomaz2022dipolar,bilinskaya2024realizing}, trapped ions setups~\cite{monroe2021programmable,wang2024realizing}, and cold atoms in cavity experiments~\cite{ritsch2013cold,mivehvar2021cavity,wu2023signatures}, making them promising candidates for building efficient quantum computing devices~\cite{ladd2010quantum,ramette2022any}.
	
	The non-interacting quasiperiodic model, best known as the Aubry‑André model~\cite{aubry1980analyticity}, has been studied for decades~\cite{larcher2011localization,das2019realizing} and realized in ultracold atomic systems via on‑site quasiperiodic potentials~\cite{roati2008anderson,schreiber2015observation}. In contrast, where the Aubry‑André model involves single‑particle dynamics (ballistic expansion or localization)~\cite{mastropietro2015localization,roosz2024front}, here we study a two‑fermion system with quasiperiodic long‑range interactions and uncover a novel correlated dynamical regime.
	
	In this work, we investigate the dynamics of two identical spinless fermions on a one-dimensional lattice with open boundary conditions (OBC), subject to quasiperiodic long-range interactions. When interactions are sufficiently strong, we uncover a correlated dynamical regime characterized by two particles walking together while maintaining an approximately constant inter-particle distance. By tuning the phase, we observe three distinct manifestations: localization (a special case of the constant-distance walking), and two deviations from it---nearest-neighbor separation oscillations and next-nearest-neighbor separation transitions.
	
	Furthermore, we identify the suppression of entanglement entropy in the system, including instances of oscillatory behavior. 
	
	\section{Model}
	We consider fermions in a lattice with quasiperiodic long-range interaction,
	\begin{equation}
		\hat{H} = -J \sum_{i=1}^{L} \left( \hat{c}_{i}^{\dagger} \hat{c}_{i+1} + \mathrm{H.c.} \right) 
		+ \sum_{1 \leq i < j \leq L} U_{ij} \hat{n}_i \hat{n}_j .
		\label{eq:H}
	\end{equation}
	Here,  $ \hat{c}_{i}^{\dagger} $  ( $ \hat{c}_{i} $ ) and  $ \hat{n}_i = \hat{c}_{i}^{\dagger}\hat{c}_{i} $  are the operators of fermionic creation (annihilation) and particle number at the  $ i $th site, respectively. The system size is $ L $.  $ J $  denotes the nearest-neighbor hopping strength. The interaction term, with $U_{ij}=\Delta \cos\left(2\pi\beta|i-j| + \phi\right)$,
	describes a quasiperiodically modulated long-range interaction between a pair of sites  $ (i,j) $ , where  $ \Delta $ ,  $ \beta $ , and  $ \phi $  represent the modulation amplitude, spatial frequency, and initial phase, respectively. Although the interaction is quasiperiodically modulated, it depends only on the relative distance \(r=|i-j|\), so the Hamiltonian is translationally invariant. The interaction strength varies with \(r\) in a cosine form with spatial period controlled by the dimensionless parameter  $ \beta $  (chosen as an irrational number to realize quasiperiodicity). The role of quasiperiodicity is further examined in the Appendix~\ref{appA} by comparing the dynamics for rational $ \beta $ and fully disordered interactions. This model can be effectively realized in experimental platforms such as superconducting circuits~\cite{tao2023interaction,li2023observation,solfanelli2024stabilization,huang2026observation} and ultracold atoms~\cite{nicholson2015optical,clark2015quantum,wang2022observation} by mapping its low-dimensional, complex interaction profiles onto an experimentally feasible higher-dimensional synthetic lattice system, where the interactions are effectively represented as simple on-site potentials~\cite{guo2025vacuum}.
	
	\section{Two-particle correlated dynamics}
	\begin{figure}[tbp]
		\includegraphics[width=0.48\textwidth]{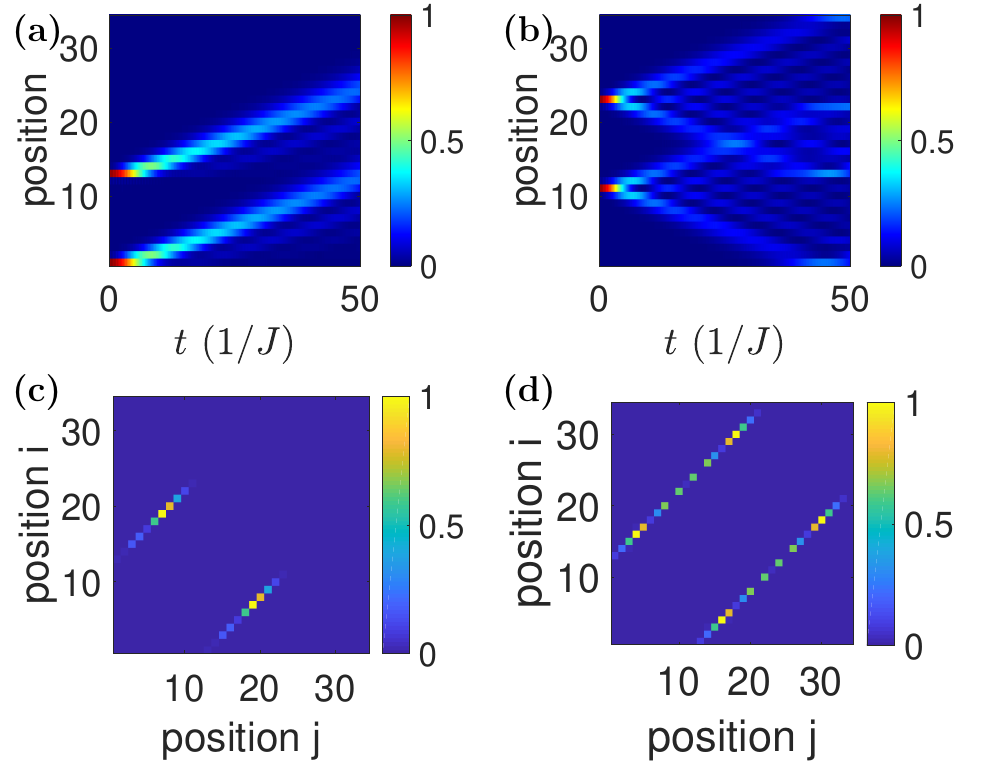}
		\caption{Probability density evolution and two-particle correlations. 
			(a, b) Probability density $P(i,t)$ (Eq.~\eqref{eq:prob_density}) up to time   $ t = 50J^{-1} $  for an initial inter-particle distance of 12, 
			with (a) one particle initially at the boundary and (b) both particles away from the boundary. 
			(c, d) Corresponding two-particle correlation functions $\Gamma(i,j)$ (Eq.~\eqref{eq:gamma}) at  $ t = 30J^{-1} $  for the same initial configurations as (a) and (b). 
			$ L = 34 $, $ \Delta = 10 $, $ \phi = 0 $. }
		\label{fig1}
	\end{figure}	
	
	Our study focuses on a continuous-time quantum walk of two interacting particles in the strong-interaction regime. We set $\hbar = 1$ throughout this work.
	The dynamics studied here is an out-of-equilibrium quantum quench: the initial state is a Fock state with a well-defined particle separation, which is not an eigenstate of the Hamiltonian, and we evolve it unitarily for $t>0$~\cite{essler2016quench,calabrese2016quantum, gibbins2024quench}. Without loss of generality, setting $ J = 1 $ and $\beta = (\sqrt{5} - 1)/2$. We observe that two identical spinless fermions, subject to quasiperiodic long-range interactions on a one-dimensional lattice with OBC, maintain an approximately constant inter-particle distance during their quantum walk, regardless of their initial separation and whether one or both particles start from the boundaries; see Fig.~\ref{fig1}.
	
	Under these conditions and for strong interactions ($\Delta = 10$), Figs.~\ref{fig1}(a) and (b) display the evolution of the single-particle probability density distribution~\cite{lahini2012quantum,wang2015quantum}:
	\begin{equation}
		P(i,t) = \langle \Psi(t) | \hat{n}_i | \Psi(t) \rangle.
		\label{eq:prob_density}
	\end{equation}
	This quantity represents the probability of finding a particle at site $i$ at time $t$. In Fig.~\ref{fig1}(a), when one particle is initially placed at the boundary, the pair is observed to walk predominantly in one direction, with the inter-particle distance remaining approximately  constant over an extended period. This asymmetry originates from the OBC: with one particle initialized at the boundary, propagation toward the boundary is blocked, so the pair can only move away from it.
	In Fig.~\ref{fig1}(b) where both particles are initially located within the bulk of the lattice chain, spatial symmetry is restored since they are free to propagate toward either edge of the chain, making the constant-distance behavior less apparent than in the boundary-initiated case in Fig.~\ref{fig1}(a). The linear light cone observed in both Figs.~\ref{fig1}(a) (half-cone) and (b) (full cone) originates from the short-range hopping term, which provides a finite propagation speed analogous to the mechanism of the Lieb-Robinson bound~\cite{lieb1972finite,kliesch2014lieb,tran2020hierarchy}.
	
	As a complementary view, Figs.~\ref{fig1}(c) and (d) show the two-particle density-density correlation function at a fixed time $t=30J^{-1}$. The correlation function is defined as
	\begin{equation}
		\Gamma(i,j,t) = \langle \Psi(t) | \hat{n}_i \hat{n}_j | \Psi(t) \rangle, \quad i \ne j,
		\label{eq:gamma}
	\end{equation}
	which quantifies the joint probability of finding one particle at site $i$ and the other at site $j$~\cite{lahini2012quantum,wang2015quantum}. We observe that $\Gamma(i,j,t) \approx \Gamma(|i-j|,t)$, i.e., it depends predominantly on the relative distance, as is already visible in Figs.~\ref{fig1}(c) and (d). The absence of the diagonal \(i=j\) in the correlation panels reflects the forbidden double occupation for spinless fermions. Additionally, the same boundary effect is also reflected in the correlation functions: in Fig.~\ref{fig1}(c), corresponding to the boundary-initiated case, the correlation pattern is asymmetric, while in Fig.~\ref{fig1}(d), for the bulk-initiated case, it is symmetric. These correlations reveal that, regardless of whether a particle starts at the boundary or not, the two particles maintain a nearly fixed separation, highlighting a coordinated, pair-like walking mode.
	
	\begin{figure}[tbp]
		\includegraphics[width=0.48\textwidth]{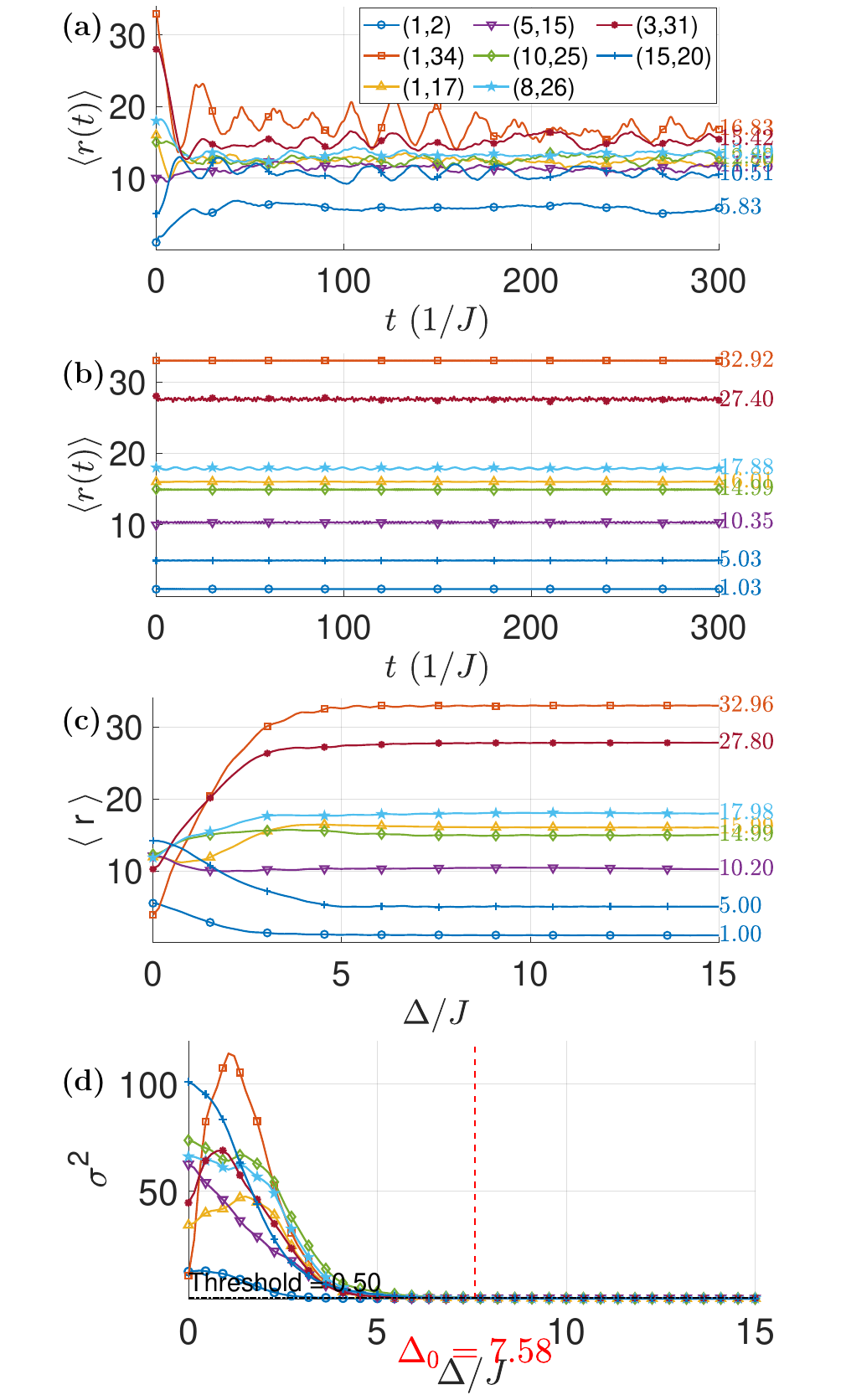}
		\caption{Expected inter-particle distance and its variance. 
			(a,b) Time evolution of the expected inter-particle distance $\langle r(t) \rangle$ (Eq.~\eqref{eq:r_avg}) up to 
			$t = 300J^{-1}$ for (a) $\Delta = 1$ and (b) $\Delta =10$. (c,d) At $t = 10J^{-1}$: (c) expected distance $\langle r \rangle$ and (d) variance (Eq.~\eqref{eq:variance}) as a function of $\Delta$.
			Dashed line in (d): variance threshold $0.50 \rightarrow \Delta_0 = 7.58$. 
			$L = 34$, $\phi = 0$.}
		\label{fig2}
	\end{figure}
	We now discuss the effects of the interaction strength $\Delta$, presented in Fig.~\ref{fig2}. As $\Delta$ increases, the inter-particle distance becomes more stable and remains close to its initial value. We consider a lattice of size $L=34$. Eight representative initial positions were selected: $(1,2),\ (1,34),\ (1,17),\ (5,15),\ (10,25),\ (8,26),\ (3,31),\\ (15,20)$. They cover a range of inter-particle distances, including cases where both particles are on the same boundary, on opposite boundaries, one on a boundary and one in the interior, or both in the interior.
	
	The expected inter-particle distance,
	\begin{equation}
		\langle r(t) \rangle = \sum_{i<j} |i - j| \, |\psi_{ij}(t)|^2,
		\label{eq:r_avg}
	\end{equation}
	where $\psi_{ij}(t) = \langle i,j | \Psi(t) \rangle$, is shown in Figs.~\ref{fig2}(a) and (b) up to $ t = 300J^{-1} $ for weak ($\Delta = 1$) and strong ($\Delta = 10$) interaction regimes. Under weak interactions, the  expected distance $\langle r(t) \rangle$ fluctuates randomly without a clear trend. In contrast, under strong interactions,  $\langle r(t) \rangle$ remains close to the initial inter-particle distance $r_0=\lvert i_0 - j_0 \rvert$, suggesting the two particles form a long-lived, loosely bound pair---that is, a correlated pair whose relative motion is confined by their quasiperiodic interactions.
	
	Fig.~\ref{fig2}(c) plots the expected distance $\langle r \rangle$ as a function of the interaction strength $\Delta$ at a fixed time $ t = 10J^{-1} $. Across all eight configurations, $\langle r \rangle$  is further stabilized and becomes well-defined as $\Delta$ increases. We therefore define a \textit{characteristic interaction strength} $\Delta_0 = 7.58$ from the point in Fig.~\ref{fig2}(d) where the variance,
	\begin{equation}
		\sigma^2 = \sum_{i<j} |i - j|^2 |\psi_{ij}|^2 - \left( \sum_{i<j} |i - j| |\psi_{ij}|^2 \right)^2,
		\label{eq:variance}
	\end{equation}
	across the eight configurations falls below $0.50$.
	
	Taken together, Figs.~\ref{fig2}(b)--(d) provide a consistent and complementary picture of the emergence of the approximate constant-distance walking. Fig.~\ref{fig2}(b) demonstrates that for sufficiently large $\Delta$, the expected inter-particle distance remains close to its initial value over a long time, confirming the dynamical robustness of the correlated walking. Fig.~\ref{fig2}(c) complements this by showing that, at a fixed time, the distance converges smoothly toward $r_0$ as $\Delta$ increases. Fig.~\ref{fig2}(d) further reveals the suppression of fluctuations in the inter-particle distance, allowing us to identify a characteristic threshold $\Delta_0$. Together, these three panels establish that the emergence of the approximate constant-distance walking is not a transient effect but a genuine dynamical regime governed by the strong quasiperiodic interactions.
	
	To quantify the propagation speed of the particles, we define an effective velocity $v_{\mathrm{eff}}$ by tracking the density front $x_f(t)$. For a given propagation direction, $x_f(t)$ is defined as the point at which the integrated density from the corresponding boundary reaches a fixed threshold $\eta$ ($0<\eta<1$):
		\begin{equation}
			\int_{x_f(t)}^{L} P(x,t)\,dx = \eta
			\quad\text{or}\quad
			\int_{0}^{x_f(t)} P(x,t)\,dx = \eta,
		\end{equation}
	with $P(x,t)$ the single-particle density distribution in Eq.~\eqref{eq:prob_density}. This definition, applicable to parameter regimes where the density evolution exhibits a well-defined light-cone structure, is standard in studies of quantum many-body dynamics \cite{foss2015nearly,mezei2017entanglement,najafi2018light,tran2020hierarchy}. In practice, we take the average slope in the linear regime of the light cone, before boundary effects set in:
		\begin{equation}
			v_{\mathrm{eff}} \simeq \left|\frac{x_f(t_2) - x_f(t_1)}{t_2 - t_1}\right|.
		\end{equation}
	In the noninteracting case ($\Delta = 0$) with $L=34$ (as in Fig.~\ref{fig1}(b)), applying the above procedure with $\eta = 0.05$ and for early times $t \leq 6J^{-1}$ yields $v_{\mathrm{eff}} \approx 1.83J$, which is close to the theoretical free-fermion value $v_{\mathrm{free}} = 2J$. This deviation is due to finite-size effects and the specific choice of the parameters \(\eta\) and the time interval.
	
\begin{figure}[tbp]
			\includegraphics[width=0.48\textwidth]{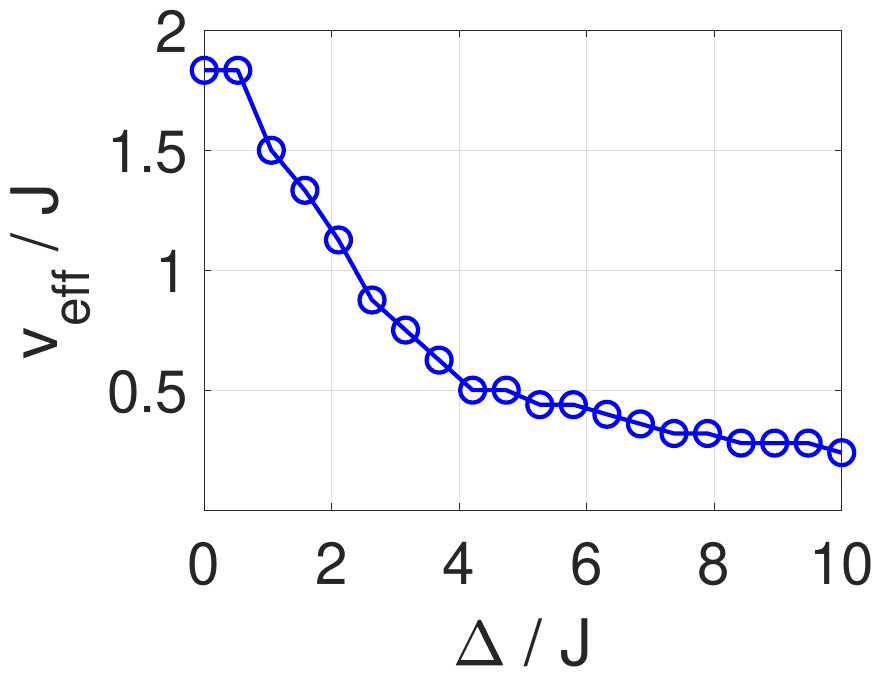}
			\caption{Effective velocity \(v_{\mathrm{eff}}\) as a function of \(\Delta\), with the same parameters as in Fig.~\ref{fig1}(b). 
			}
			\label{fig3}
		\end{figure}

	For finite interaction strengths, we apply the same extraction method using longer time intervals for stronger interactions. As shown in Fig.~\ref{fig3}, the effective velocity decreases with increasing $\Delta$. 
	
	The quasiperiodic modulation yields a sign-alternating pattern of $U_{ij}$ as the inter-particle distance changes by one lattice unit, with consecutive same signs in some cases. When interactions dominate, this structure suppresses deviations from the initial separation, maintaining an approximately constant inter-particle distance.
	
	When $\Delta/J$ is sufficiently large such that interactions dominate, we first study the eigenstates via exact diagonalization (ED). The eigenstates $|\phi_n\rangle$ are defined by the eigenvalue equation:
	\begin{equation}
		\hat{H} |\phi_n\rangle = E_n |\phi_n\rangle.
		\label{eq:EqE}
	\end{equation}
	Applying ED to our model, we find that these eigenstates are predominantly composed of Fock states with a fixed inter-particle distance. To characterize this, when the system is initialized in a specific configuration $|\Psi(0)\rangle$, the projection weight onto each eigenstate is defined as
	\begin{equation}
		|c_n|^2 = |\langle \phi_n | \Psi(0) \rangle|^2.
		\label{eq:projection_weight}
	\end{equation}
	We further define the \textit{characteristic distance} of an eigenstate $|\phi_n\rangle$ as the expectation value of the inter-particle distance within that eigenstate:
	\begin{equation}
		\bar{r}_n = \sum_{k} r_k \, |\psi_n(k)|^2,
		\label{eq:characteristic_distance}
	\end{equation}
	where the index $k$ runs over all Fock states $|i,j\rangle$ with $i<j$, $r_k = |i - j|$ for the Fock state $|i, j\rangle$ and $|\psi_n(k)|^2$ is its probability in $|\phi_n\rangle$. With this definition, we find that the projection weights $|c_n|^2$ concentrate on eigenstates whose \textit{characteristic distances} $\bar{r}_n$ lie near $r_0$. As concrete examples, for the configuration of Fig.~\ref{fig1}(a), the total weight on eigenstates with distance exactly $r_0$ is approximately $98.1\,\%$, while the cumulative weight on eigenstates with distances in $[r_0-1, r_0+1]$ exceeds $99.5\,\%$; for Fig.~\ref{fig1}(b), the corresponding weights are $97.6\,\%$ and $99.2\,\%$, respectively. Such a distribution ensures that the time evolution is governed by a set of eigenstates sharing a nearly constant inter-particle distance, giving rise to the correlated walking dynamics observed in Fig.~\ref{fig1}.
	
	Complementary to the eigenstate analysis, an energy perspective supports the observed behavior. Sign-alternating quasiperiodic interaction leads to large energy differences:
		\begin{equation}
			\Delta E_1 = |\langle E\rangle_{r + 1} - \langle E\rangle_r|,\quad
			\Delta E_2 = |\langle E\rangle_{r + 2} - \langle E\rangle_r|,
			\label{eq:delta_E}
		\end{equation}%
	where $\langle E \rangle_r$ is the energy expectation value of a configuration with separation $r$. Large $\Delta E_1$ and $\Delta E_2$ suppress single-step and double-step changes of inter-particle distance, thereby stabilizing the constant-distance walk. For our strong-interaction regime ($\Delta/J = 10$), energy is dominated by interaction energy $U$ (which depends on $r$), with absolute difference $|\langle E \rangle - U| \sim 10^{-4}$. In the following discussion (except for precise numerical fitting) we use $U$ as an approximation for $\langle E \rangle$.
	
	\section{Special regimes of correlated dynamics}
	By tuning the phase $\phi$, this phenomenon exhibits three distinct manifestations: localization (a special case of the constant-distance walking), and two deviations from it---nearest-neighbor separation oscillations and next-nearest-neighbor separation transitions---each arising for specific initial separations $r_0$.
	
\begin{figure}[tbp]
		\includegraphics[width=0.48\textwidth]{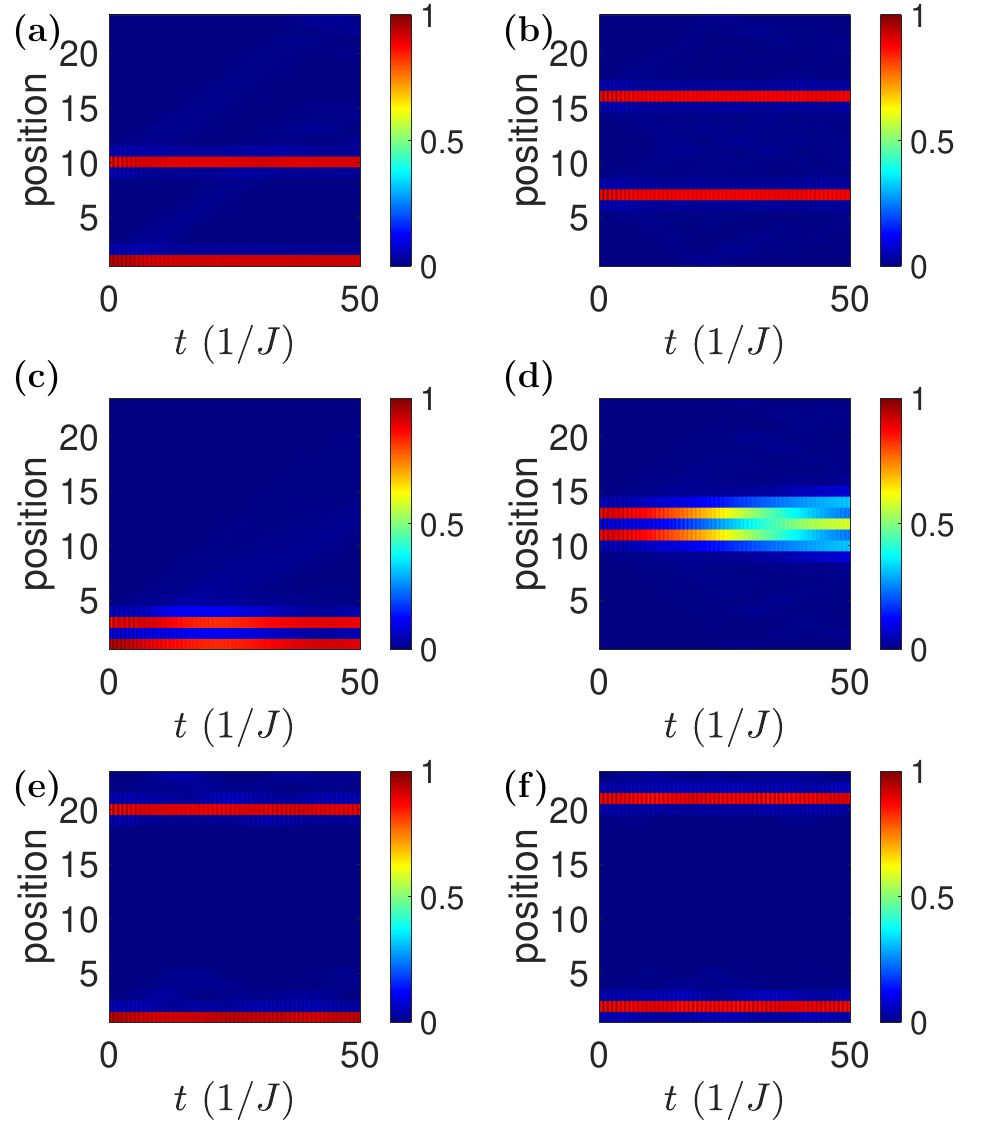}
		\caption{Localization dynamics of two particles.
			The left (right) column shows initial configurations with one (neither) particle at the boundary. 
			(a,b) For phase $\phi = 3\pi/8$ and initial separation $9$. 
			(c--f) For the fixed phase $\phi = \pi/64$, with two distinct initial separations: 
			(c,d) separation $= 2$; (e,f) separation $= 19$. 
			$L = 23$, $\Delta = 10$.}
		\label{fig4}
	\end{figure}

	First, as shown in Fig.~\ref{fig4}, we consider a lattice of size $L=23$ and an evolution time $t=50J^{-1}$. The probability density evolution (Eq.~\eqref{eq:prob_density}) reveals that, depending on the phase $\phi$, the two particles exhibit long-time localization  for one or more specific initial distances. This occurs when $U$ is close to zero for those distances. This is because the eigenstates onto which the initial state projects have energies concentrated near zero, and are nearly degenerate. Such near-degeneracy enhances the role of interference terms in the time evolution.
	
	For the density distribution $P(i, t)$ (Eq.~\eqref{eq:prob_density}), the expansion in the eigenstate basis yields~\cite{sakurai2020modern}:
	\begin{multline}
		P(i, t) = \sum_{n} |c_n|^2 \langle \phi_n | \hat{n}_i | \phi_n \rangle \\
		+\sum_{m \neq n} c_m^* c_n \langle \phi_m | \hat{n}_i | \phi_n \rangle e^{i(E_m - E_n)t},
		\label{eq:prob_density_expansion}
	\end{multline}
	where the first term is the weighted average of the eigenstate densities, and the second term encodes quantum interference between different eigenstates. This interference can lead to localization even when the participating eigenstates are individually extended---a phenomenon known as interference localization, which plays a central role in dynamical localization~\cite{muller2010disorder}.
	
	For $\phi = 3\pi/8$, with initial separation $r_0=9$ (Figs.~\ref{fig4}(a) and (b)), the boundary initial state $|1, 10\rangle$ projects predominantly onto a single strongly localized eigenstate (inverse participation ratio IPR $\approx 0.888$, where $\mathrm{IPR} = \sum_k |\psi_n(k)|^4$ and approaches 1 for a fully localized state), directly yielding localization as seen in Fig.~\ref{fig4}(a). In contrast, the interior initial state $|7,16\rangle$ projects onto many eigenstates with nearly degenerate energies near zero, the eigenstate with the largest weight being extended (IPR $\approx 0.129$). The coherent superposition of these eigenstates, via the interference terms in Eq.~\eqref{eq:prob_density_expansion}, produces the confined wave packet observed in Fig.~\ref{fig4}(b).
	
	For $\phi = \pi/64$ (Figs.~\ref{fig4}(c)–(f)), there are two initial distances leading to localization: $2$ and $19$. Notably, at distance $19$ the interaction $U$ is closer to zero, resulting in stronger localization. As seen in Figs.~\ref{fig4}(c) and (d), with all other conditions the same, localization is stronger at the boundary than in the interior. Unlike the boundary case in Fig.~\ref{fig4}(a), where the initial state projects predominantly onto a single strongly localized eigenstate (IPR $\approx 0.888$), the localization in Figs.~\ref{fig4}(c) and (e) involves two eigenstates (IPR $\approx 0.44–0.45$), while the interior cases in Figs.~\ref{fig4}(d) and (f) arise from the coherent superposition of two or more near‑degenerate eigenstates.
	
	
	Second, in specific cases, projection onto two distinct distance sectors (nearest-neighbor or next-nearest-neighbor) yields oscillatory or transitional dynamics, respectively: oscillations when $\Delta E_1$ is small, transitions when $\Delta E_2$ is negligible. See below for details.
	
\begin{figure}[tbp]
		\includegraphics[width=0.48\textwidth]{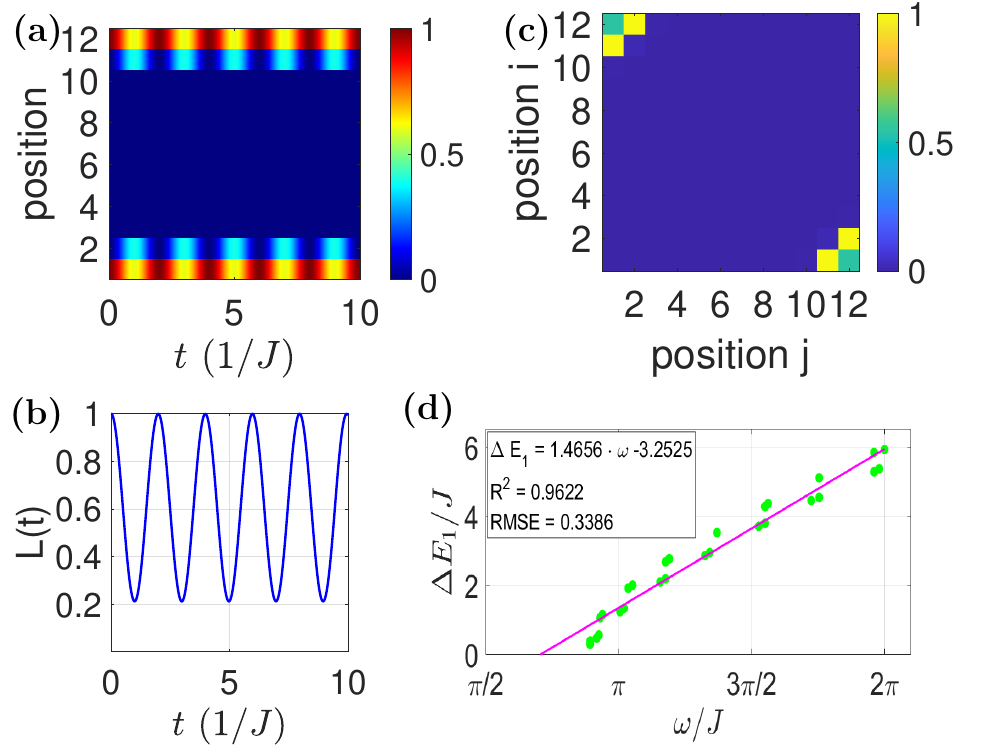}
		\caption{Nearest-neighbor separation oscillations at the boundaries. 
			Evolution of (a) the probability density distribution and (b) the Loschmidt echo (Eq.~\eqref{eq:Loschmidt_echo}) up to $t = 10J^{-1}$. 
			(c) Correlation function at the minimum of $L(t)$ ($t=1J^{-1}$). 
			$L = 12$, $\Delta = 10$, $\phi = 0$. 
			(d) Angular frequency $\omega$ of $L(t)$ oscillation versus $\Delta E_1$. The 28 data points are selected from lattices of size $L = 8$--$71$ with $\Delta = 10$, $\phi = 0$ or $\pi/2$, where oscillations are pronounced ($\omega \in (0, 2\pi]$, corresponding to period $T \ge 1$).}
		\label{fig5}
	\end{figure}

	The first of these, oscillations, is isolated by considering the boundary configuration (particles at opposite ends). Under specific conditions, this configuration exhibits oscillatory dynamics rather than the typical boundary localization. We examine the initial state's projection onto eigenstates for the parameters of Fig.~\ref{fig5} ($L=12$, $\Delta = 10$, $\phi = 0$, initial state $|1, 12\rangle$, $r_0=11$). ED reveals that the total weight on eigenstates with \textit{characteristic distances} exactly $r_0$ is $72.98\,\%$, while the cumulative weight on eigenstates with distances in $[10,11]$ reaches $99.98\,\%$. Thus, the initial state projects predominantly onto the two distance sectors $r_0$ and $r_0-1$, which enables the oscillatory dynamics. 
	
	The evolution of the probability density in Fig.~\ref{fig5}(a) illustrates this oscillation. The Loschmidt echo (also known as the quantum return probability for pure states)~\cite{adamov2003loschmidt,krapivsky2018quantum,campos2010unitary} in Fig.~\ref{fig5}(b) is defined as
	\begin{equation} 
		L(t) = \left| \langle \Psi(0) | e^{-i\hat{H}t} | \Psi(0) \rangle \right|^2,
		\label{eq:Loschmidt_echo} 
	\end{equation}
	which quantifies the similarity between the time-evolved state and the initial state. It equals 1 at $t=0$ and approaches 0 when the state is nearly orthogonal. Fig.~\ref{fig5}(c) shows the two-particle correlation function at the minimum of $L(t)$ ($t=1J^{-1}$; see Fig.~\ref{fig5}(b)), indicating that the inter-particle distance is most likely one lattice unit less than the initial separation. We have also checked the correlation function at later local minima of the Loschmidt echo, e.g., \(t=9J^{-1}\), and found qualitatively the same pattern, confirming that the oscillation persists and that the key physics is already captured at the first minimum.
	
	Expanding the initial state in the eigenbasis of $\hat{H}$ (Eq.~\eqref{eq:EqE}), Eq.~\eqref{eq:Loschmidt_echo} can be rewritten as:
	\begin{equation} 
		L(t) = \sum_{n} p_n^2 + 2 \sum_{n < m} p_n p_m \cos\left( (E_n - E_m)t \right).
		\label{eq:Loschmidt echo_2} 
	\end{equation}
	Here, $p_n = |c_n|^2$. The oscillatory terms in $L(t)$ are a superposition of cosine functions. In the two-state case, the angular frequency $\omega$ of the Loschmidt echo is simply the energy difference: $\omega_{nm} = E_n - E_m$. However, the system behavior is more complex than an ideal two-level model, so we cannot directly apply this simple relation.We propose a phenomenological linear relation:
	\[
	\Delta E_1 = a \cdot \omega + b,
	\]
	where $a$ and $b$ are fitting parameters.
	To test this hypothesis, we select $28$ sample data points from $L = 8$ to $71$ with $\phi = 0$ or $\pi/2$, where boundary oscillations are pronounced. A least-squares fit of $\omega$ against $\Delta E_1$ (Fig.~\ref{fig5}(d)) yields a good linear relation:
	\[
	\Delta E_1 = 1.4656 \, \omega - 3.2525,
	\]
	with a coefficient of determination $R^2 = 0.9622$ and a root-mean-square error $\text{RMSE} = 0.3386$. This linear relation is only an empirical description valid within the sampled frequency range; according to Eq.~\eqref{eq:delta_E}, \(\Delta E_1\) is non-negative, which imposes a positive lower bound on \(\omega\) and makes the fit inapplicable in the small-\(\Delta E_1\) regime; moreover, the RMSE indicates that the linear fit works better for large values of \(\Delta E_1\) than for small ones.
	
	\begin{figure}[tbp]
		\includegraphics[width=0.48\textwidth]{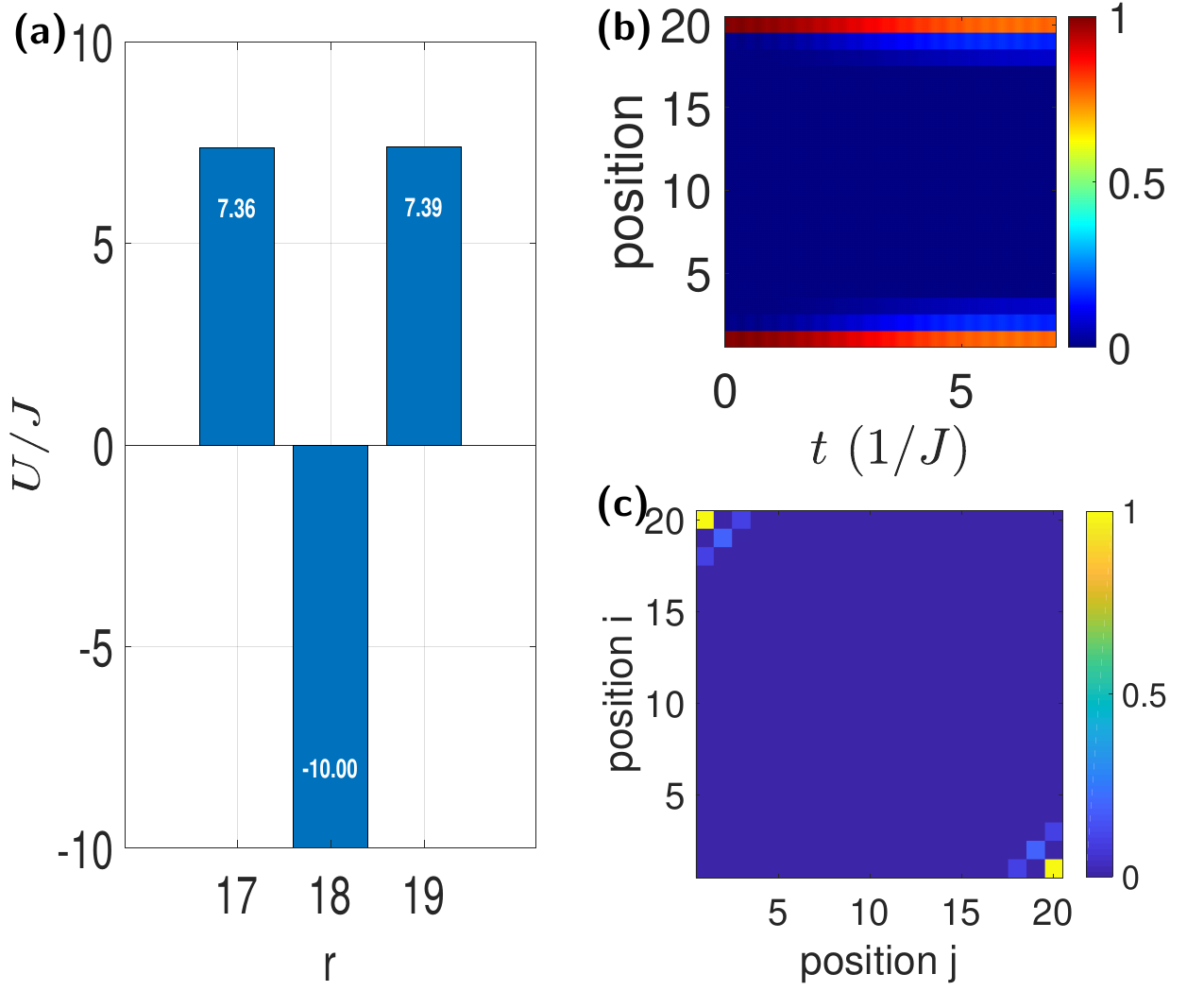}
		\caption{Next-nearest-neighbor separation transitions. 
			(a) Interaction energies for initial separations of $17$, $18$, and $19$. 
			(b) Time evolution of the probability density (two particles initially at boundaries) up to $t = 7J^{-1}$. 
			(c) Correlation function at $t = 7J^{-1}$. $L = 20$, $\Delta = 10$, $\phi = 3\pi/4$.}
		\label{fig6}
	\end{figure}
	
	The second of these, transitions, is realized by precisely tuning the phase $\phi$. We can always find cases where the interaction signs are opposite for particle separations $r$ differing by one lattice unit, while the interaction energies $U$ are nearly equal for separations $r$ differing by two lattice units, i.e., $U(r) \approx U(r + 2)$, as shown in Fig.~\ref{fig6}(a). Because $\phi$ is a continuous parameter, one can achieve exact equality $U(r) = U(r+2)$ (and thus $\Delta E_2 = 0$) by solving $\cos(2\pi\beta r + \phi) = \cos(2\pi\beta(r+2) + \phi)$. With $r = 17$ and $\beta = (\sqrt{5} - 1)/2$, one obtains $\phi \approx 0.751\pi$. For simplicity, take $\phi = 3\pi/4$. Consider separations differing by one unit, e.g., $r=18$ and $19$: $U(18) = -10$ and $U(19) = 7.39$, resulting in a large energy difference $\Delta E_1 \approx |U(19) - U(18)| = 17.39$. Such a large energy barrier would keep  the distance constant. However, for separations differing by two lattice units, e.g., $r = 17$ and $r = 19$, the interaction energies are $U(17) = 7.36$ and $U(19) = 7.39$, respectively, yielding a negligibly small $\Delta E_2 \approx 0.03$.  
	
	For the parameters of  Figs.~\ref{fig6}(b) and (c) ($L=20, \Delta=10, \phi=3\pi/4$, initial separation $r_0=19$), ED reveals that the total weight on eigenstates with \textit{characteristic distances} $r_0=19$ is $78.90\,\%$, while the weight on eigenstates with distance $r_0-2=17$ is $20.46\,\%$; the cumulative weight on these two distance sectors exceeds $99\,\%$. This yields next‑nearest‑neighbor separation transitions: with a small but finite probability of changing distance by two units (from 19 to 17) during evolution. Thus, despite the large $\Delta E_1$ suppressing single-step changes, the small $\Delta E_2$ allows occasional double-step transitions, demonstrating that $\Delta E_2$ plays a significant role in the approximate constant-distance walk.
	
	
	\begin{figure*}[tbp]
		\includegraphics[width=0.98\textwidth]{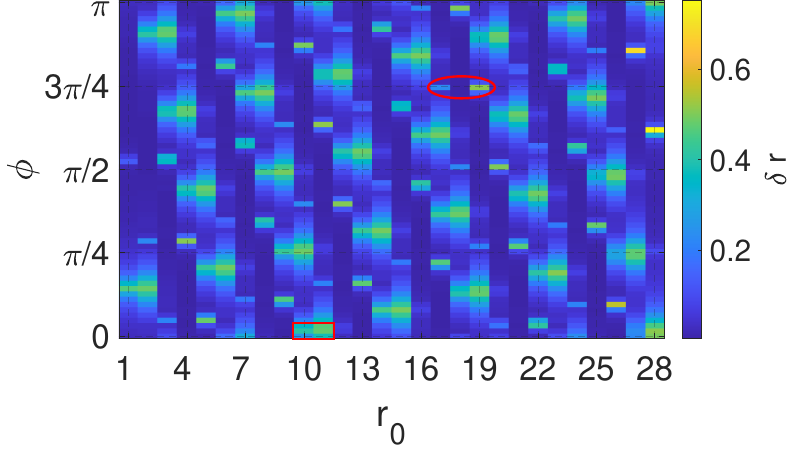}
		\caption{Phase diagram. The time-averaged deviation $\delta r$ (Eq.~\eqref{eq:delta_r})  for $L = 29$, $\Delta = 10$, $\beta = (\sqrt{5} - 1)/2$, and $T = 10J^{-1}$. Rectangular red box: nearest-neighbor oscillations (Fig.~\ref{fig5}); elliptical red box: next-nearest-neighbor transition (Fig.~\ref{fig6}).
		}
		\label{fig7}
	\end{figure*}
	
	\section{Phase diagram}
	To present a global picture of the approximate constant distance walking (with localization as one of its manifestations) as well as the two distinct regimes (nearest-neighbor oscillations and next-nearest-neighbor transitions), Fig.~\ref{fig7} shows a phase diagram in $(r_0, \phi)$ space of the time-averaged deviation:
	\begin{equation}
		\delta r = \frac{1}{T} \int_{0}^{T} \left| \langle r(t) \rangle - r_0 \right| dt,
		\label{eq:delta_r}
	\end{equation}
	for $L = 29$, $\Delta = 10$, $\beta = (\sqrt{5} - 1)/2$, and $T = 10J^{-1}$.
	
	The phase diagram reveals a quasiperiodic pattern in $\delta r$ as a function of the initial separation $r_0$ and the phase $\phi$, originating from the incommensurate spatial frequency $\beta$ in the cosine-modulated long-range interaction. For most $(r_0, \phi)$ pairs, $\delta r$ remains small (blue regions), indicating robust constant-distance walking. Specific combinations, e.g., the rectangular red box at $\phi = 0$, $r_0 = 10, 11$ (nearest-neighbor oscillations, Fig.~\ref{fig5}) and the elliptical red box at $(r_0 = 19, \phi = 3\pi/4)$ (next-nearest-neighbor transition, Fig.~\ref{fig6}) exhibit clear deviations. The full distribution of such deviations across $(r_0, \phi)$ reveals the quasiperiodic nature.
	
	We verified that boundary reflections do not alter this behavior; after reflection, two particles continue to maintain an approximately constant-distance walk.
	
	\section{Entanglement entropy}
	For a two-particle system, the entanglement entropy is defined as the von Neumann entropy of the single-particle reduced density matrix $\rho_1(t)$~\cite{affleck2009entanglement,ferreira2022quantum,zhong2025quantum}:
	\begin{equation}
		S_1(t) = -\mathrm{Tr}[\rho_1(t)\log_2 \rho_1(t)],
		\label{eq:S(t)}
	\end{equation}
	where the matrix elements of the single-particle reduced density matrix are given by:
	\begin{equation}
		[\rho_1(t)]_{x,x'} = \langle\Psi(t)|\hat{c}_x^\dagger \hat{c}_{x'}|\Psi(t)\rangle,
		\label{eq:rho_1(t)}
	\end{equation}
	with $x,x'=1,2,\dots,L$ labeling the lattice sites, and $|\Psi(t)\rangle$ being the two-particle state vector (consistent with the definitions in Eqs. \eqref{eq:prob_density} and \eqref{eq:gamma}).
	
	\begin{figure}[tbp]
		\includegraphics[width=0.48\textwidth]{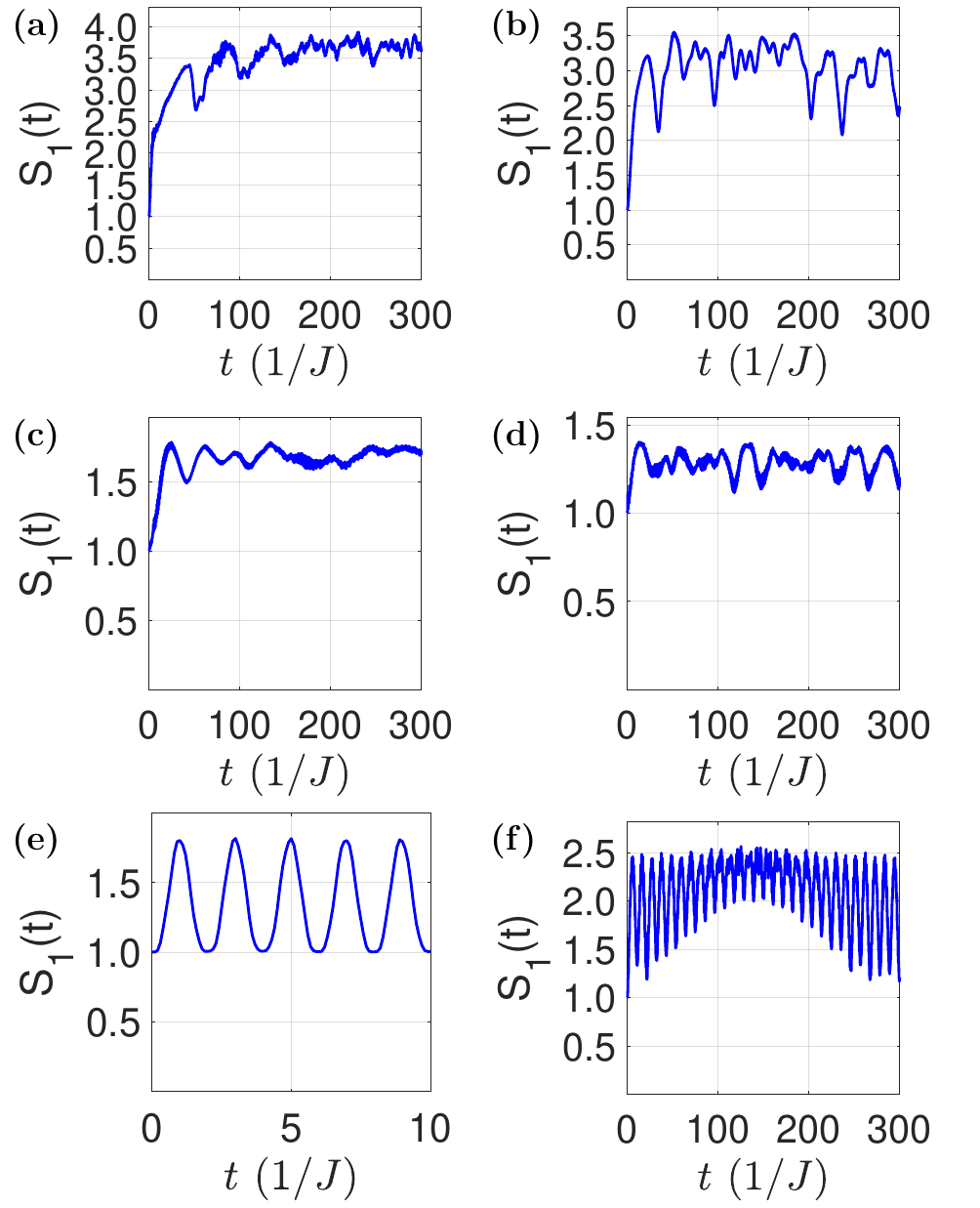}
		\caption{Entanglement entropy (Eq.~\eqref{eq:S(t)}).
			(a) Initial separation $3$ and (b) initial separation $17$, with $L = 23$, $\Delta = 10$, $\phi = 0$, $t = 300J^{-1}$. 
			(c) Initial separation $2$ (cf. Fig.~\ref{fig4}(c)) and (d) initial separation $19$ (cf. Fig.~\ref{fig4}(e)), with $L = 23$, $\Delta = 10$, $\phi = \pi/64$, $t = 300J^{-1}$. 
			(e) Oscillation at the boundary; $L = 12$, $\Delta = 10$, $\phi = 0$, $t = 10J^{-1}$. 
			(f) Initial positions $(1,\,21)$; same parameters as in (a).
		}
		\label{fig8}
	\end{figure}
	
	For our system with $L = 23$ lattice sites (Figs.~\ref{fig8}(a)–(d) and (f)) and  $L = 12$ lattice sites (Fig.~\ref{fig8}(e)), the maximum of the single-particle reduced von Neumann entropy is $\log_2 23 \approx 4.52$ and $\log_2 12 \approx 3.59$, respectively, which are attained only when the reduced density matrix is maximally mixed. We identify a clear suppression of entanglement entropy in all panels of Fig.~\ref{fig8}.
	
	Comparing Figs.~\ref{fig8}(a) and (b), the suppression of entanglement entropy is more pronounced for larger particle separations. Figs.~\ref{fig8}(c) and (d) correspond to Figs.~\ref{fig4}(c) and (e), indicating that in the localized regime, the entanglement entropy remains relatively stable over long times. A higher degree of localization (e.g., Fig.~\ref{fig4}(e)) corresponds to lower entanglement entropy, and vice versa. Fig.~\ref{fig8}(e) illustrates boundary oscillations. In this scenario, the entanglement entropy also oscillates, consistent with the oscillatory behavior of $L(t)$  (i.e., the same frequency but different amplitude) in Fig.~\ref{fig5}(b). Fig.~\ref{fig8}(f) shows a special oscillatory case ($L = 23$, initial positions $(1, 21)$); the entanglement entropy exhibits an unusual oscillatory pattern, also consistent with $L(t)$.
	
	\section{Extension to three particles}
	To gain insight into the behavior of more than two particles, we perform a preliminary numerical check for three spinless fermions under the same parameter settings as in the two-particle case. We choose four representative initial positions: $(1,13,25)$, $(7,11,16)$, $(1,6,12)$, and $(1,10,20)$, corresponding to those in Figs.~\ref{fig1}(a), \ref{fig4}(b), \ref{fig5}(a), and \ref{fig6}(b), respectively.
		\begin{figure}[tbp]
			\includegraphics[width=0.48\textwidth]{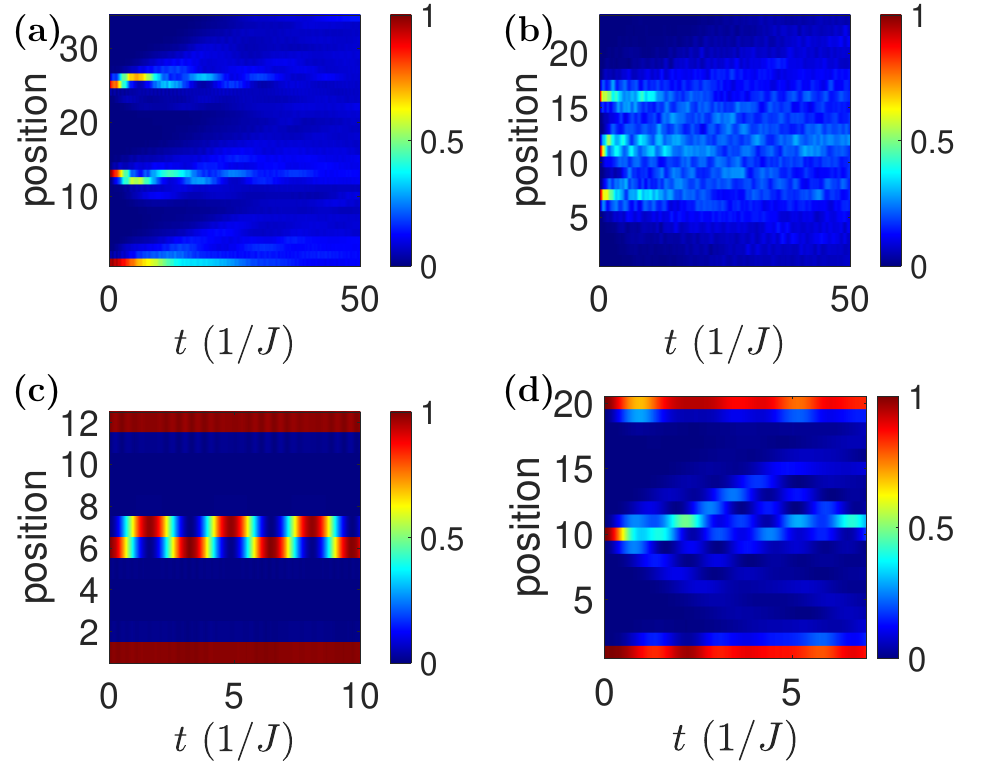}
			\caption{Probability density evolution for three spinless fermions. (a)--(d) correspond to the same parameter settings as in Figs.~\ref{fig1}(a), \ref{fig4}(b), \ref{fig5}(a), and \ref{fig6}(b), with initial positions $(1,13,25)$, $(7,11,16)$, $(1,6,12)$, and $(1,10,20)$, respectively.
			}
			\label{fig9}
		\end{figure}	
	As shown in Fig.~\ref{fig9}, the three-particle dynamics no longer exhibits the approximate constant-distance walking observed in the two-particle case. This is because the interaction now involves multiple relative distances rather than a single one, making the dynamics more complex. Only in special cases, such as those shown in Figs.~\ref{fig9}(c) and (d), does the system retain some partial features of the correlated motion.
	
	\section{Conclusion and outlook}
	A phenomenon of approximately constant-distance walk is observed for two particles in a one-dimensional lattice under strong quasiperiodic long-range interactions. The physical origin of this phenomenon lies in the translational invariance of the Hamiltonian, which stems from the interaction depending only on the relative distance between particles. This invariance, together with the energy penalty for changing the particle separation, suppresses large deviations from the initial distance and gives rise to the correlated pair dynamics. Among the three manifestations we have focused on, localization---where the inter-particle distance remains fixed---represents a special form of this constant-distance walk. The other two manifestations, nearest-neighbor separation oscillations and next-nearest-neighbor separation transitions, account for the approximate nature of the walk. Although our study of the two-particle case is limited to OBC and fermions, the mechanism suggests that this phenomenon can also appear in infinite chains and for hard-core bosons. Our work extends understanding of dynamics in few-body models with long-range interactions and quasiperiodic elements. Furthermore, the investigation of entanglement entropy in various scenarios of this model may advance related fields in quantum information. Our findings open up the possibility of studying similar long-range lattice models. Future directions include the modulation of the spatial frequency of the quasiperiodic interactions and its effect on these phenomena.
	
	\begin{acknowledgments}
			We thank Tian-Shu Deng and Yan-Xie Liu for careful guidance and invaluable discussions. We also thank the referees for their constructive comments that helped improve this manuscript.
	\end{acknowledgments}

	\bibliography{refs}
	
	\appendix
	\section{Dynamics for Rational $\beta$ and Fully Disordered Interactions}
	\label{appA}
	
	\begin{figure}[tbp]
		\includegraphics[width=0.48\textwidth]{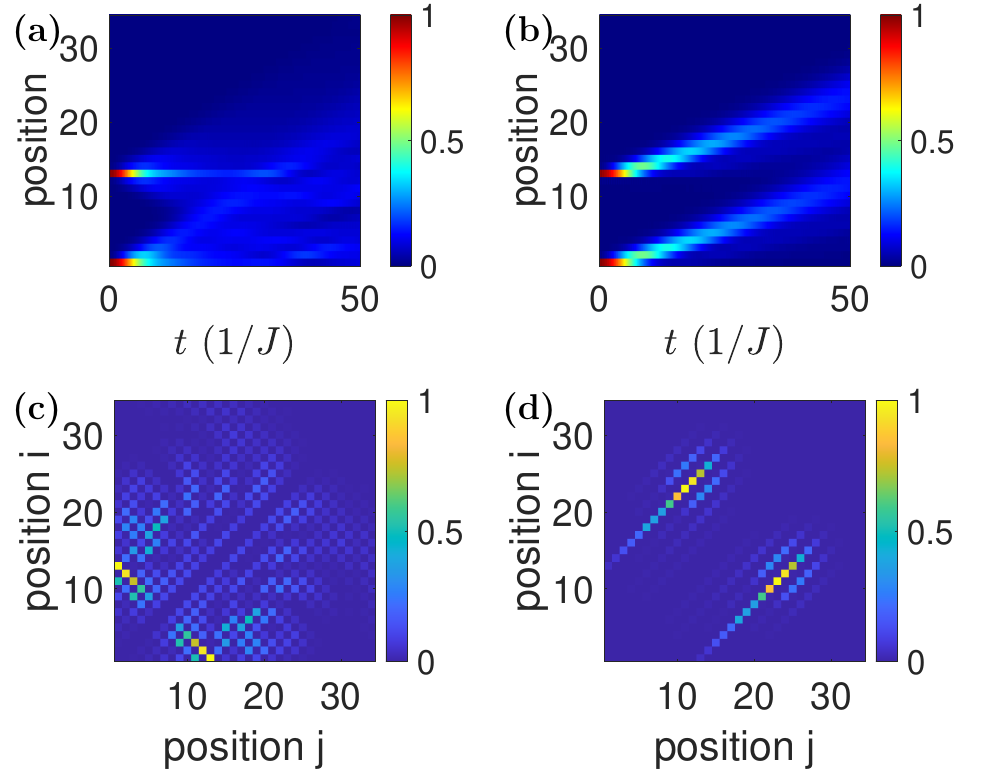}
		\caption{Probability density evolution and two-particle correlations for rational $\beta$ values. 
			(a, b) Probability density $P(i,t)$ up to time $t = 50J^{-1}$ for initial configurations (1,13) with $\beta = 1/2$ in (a) and $\beta = 2/3$ in (b). 
			(c, d) Corresponding two-particle correlation functions $\Gamma(i,j)$ at $t = 50J^{-1}$ for the same initial configurations  and same $\beta$ values as (a) and (b), respectively. 
			$L = 34$, $\Delta = 10$, $\phi = 0$.}
		\label{fig10}
	\end{figure}	
	
	In numerical simulations, a true irrational number cannot be realized exactly; we use high-precision floating-point numbers as rational approximants with extremely large denominators ($\sim 10^{15}$). Within finite system sizes and finite times, the physical behavior is indistinguishable from that of an ideal quasiperiodic system. For comparison, we also consider simple rational numbers with small denominators (e.g., $\beta=1/2$ and $2/3$), for which the interaction exhibits strictly short periods (periods 2 and 3, respectively). This leads to dynamical behaviors that are drastically different from the approximately constant-distance walking described in the main text (see  Fig.~\ref{fig10}). This comparison highlights the crucial role of quasiperiodic (irrational) modulation.

		\begin{figure}[bp]
		\includegraphics[width=0.48\textwidth]{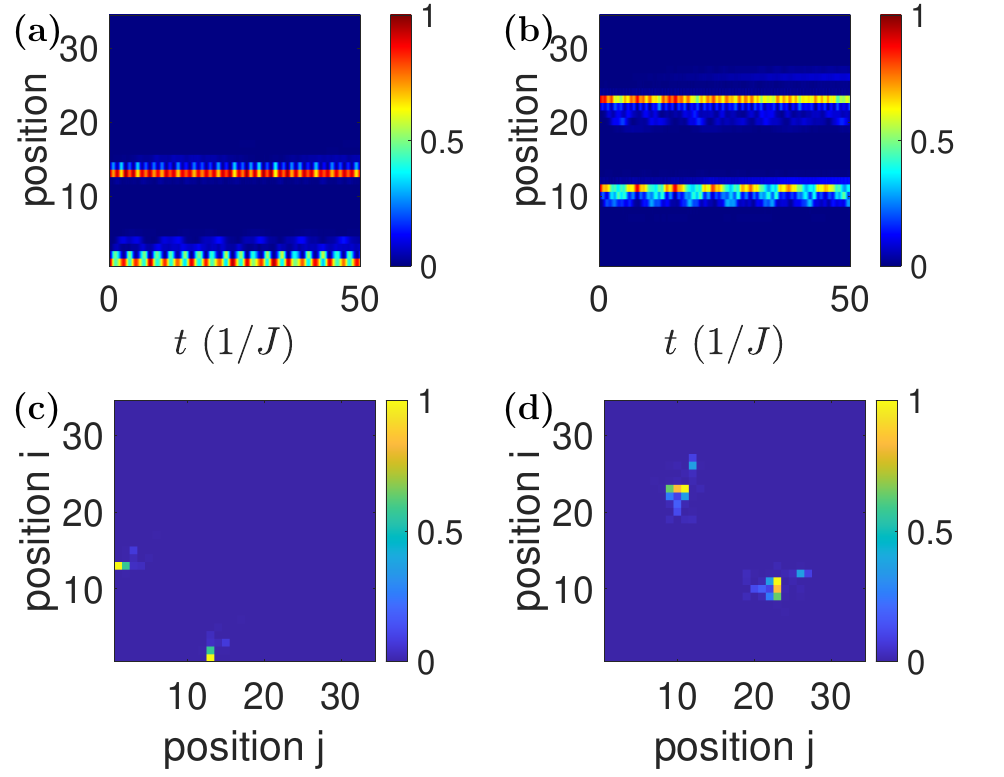}
		\caption{Probability density evolution and two-particle correlations for fully disordered interactions. 
			(a, b) Probability density $P(i,t)$ up to time $t = 50J^{-1}$ for initial configurations (1,13) in (a) and (11,23) in (b). 
			(c, d) Corresponding two-particle correlation functions $\Gamma(i,j)$ at $t = 50J^{-1}$ for the same initial configurations as (a) and (b), respectively. 
			$L = 34$, $\Delta = 10$, and each $U_{ij} \sim \text{Uniform}(-\Delta, +\Delta)$.}
		\label{fig11}
	\end{figure}	
	
	Additionally, although a systematic analysis is not performed here, we observe that the case $\beta = 2/3$ yields dynamics more reminiscent of the quasiperiodic regime than $\beta = 1/2$, suggesting that increasing the denominator of the rational approximant brings the behavior closer to the irrational limit.
	
	In the fully disordered case, we consider the Hamiltonian:
	\[
	\hat{H} = -J\sum_{i=1}^{L}\left(\hat{c}_i^{\dagger}\hat{c}_{i+1} + \mathrm{H.c.}\right) + \sum_{1\leq i<j\leq L}U_{ij}\hat{n}_i\hat{n}_j, \tag{A1}
	\]
	where the hopping term remains unchanged (still nearest-neighbor hopping $J$), but the interaction term no longer depends on the inter-particle distance $r = |i - j|$. Instead, each pair of sites $(i, j)$ independently takes a random number: $U_{ij} \sim \text{Uniform}(-\Delta, +\Delta)$, and the $U_{ij}$ for different $(i, j)$ are independent of each other and independent of the distance $r$.
	
	From Fig.~\ref{fig11}, the particles exhibit disorder-induced localization (analogous to Anderson localization), which is different from the approximately constant-distance walking dynamics described in the main text. This analogy stems from the fact that the random interactions $U_{ij}$ break translational invariance and induce destructive interference of the two-particle wavefunction, suppressing diffusion---similar to the effect of a random on-site potential in single-particle Anderson localization. The results shown in Fig.~\ref{fig11} are obtained from a single realization of fully disordered interactions (with a fixed random seed). Similar behavior is observed for other realizations.

\end{document}